\documentclass[conference]{IEEEtran}
\usepackage{cite}
\usepackage{amsmath,amssymb,amsfonts}
\usepackage{algorithmic}
\usepackage{graphicx}
\usepackage{textcomp}
\usepackage{xcolor}
\usepackage{svg}
\usepackage{gensymb}
\begin{document}

\title{SLASh: Simulation of LISLs Aboard\\LEO Satellite Shells}

\author{
\IEEEauthorblockN{Davy Romine}
\IEEEauthorblockA{\textit{Ohio State University} \\
romine.98@osu.edu}
\and
\IEEEauthorblockN{Andrew Kingery}
\IEEEauthorblockA{\textit{Ohio State University} \\
kingery.39@osu.edu}
\and
\IEEEauthorblockN{Guanqun Song}
\IEEEauthorblockA{\textit{Ohio State University} \\
song.2107@osu.edu}
\and
\IEEEauthorblockN{Ting Zhu}
\IEEEauthorblockA{\textit{Ohio State University} \\
zhu.3445@osu.edu}
}
\maketitle

\begin{abstract}
Recent advances in satellite technology have introduced a new frontier of wireless networking by establishing Low Earth Orbit (LEO) Satellite networks that work to connect difficult to reach areas and improve global connectivity. These novel advancements lack robust open-source simulation models that can highlight potential bottlenecks or potential wasted resources, wasting terrestrial users and the companies that provide these networks time and money. To that end, we propose SLASh, a highly-customizable satellite network simulation which allows users to design a simulated network with specific characteristics, and constructs them analog to real-world conditions. Additionally, SLASh can generate abstract telemetry that can be simulated moving throughout the network, allowing users to compare network capabilities across a variety of frameworks.
\end{abstract}

\begin{IEEEkeywords}
LEO Satellite Networks, Satellite constellations, Wireless networking, Computer simulation
\end{IEEEkeywords}

\section{Introduction}
Low Earth Orbit (LEO) satellite networks continue to progress and populate Earth’s atmosphere, including those of Starlink, OneWeb, and Shanghai Spacecom Satellite Technology (SSST). These satellite networks provide critical internet infrastructure to areas that otherwise would face extreme difficulty obtaining reliable internet connections, such as remote research sites far from population centers or rural communities that lack physical connections to terrestrial Internet networks. The proliferation of LEO satellite networks has increased dramatically over the current decade, and liberal projections for the number of satellites present in low earth orbit exceed 50,000 by 2030\cite{b7}. \par
Scientists studying orbital mechanics have growing concerns regarding the increasing population of LEO satellites orbiting Earth. These concerns stem from various potential issues regarding overpopulation within the Earth's atmosphere, including collision, pollution from defunct satellites, or interference with celestial observations\cite{b10}\cite{b11}. Additionally, the companies providing these satellites have a vested interest in ensuring their networks are as resource-efficient as possible for obvious economic motives. Despite this, there is a distinct lack of research in the public sector regarding the analysis of the efficiency of these networks, raising transparency issues regarding the true potential of each satellite. \par
As an answer to these concerns, we developed a \textit{\textbf{S}imulation of \textbf{L}ISLs \textbf{A}board LEO Satellite \textbf{Sh}ells (SLASh)} which aims to provide interested parties with an engine to explore the dynamics of inter-satellite communication. SLASh comes equipped with a robust satellite shell customization engine, allowing users to generate a satellite network to their specifications. These specifications can be used to simulate real-world orbital shells using publicly available data, or define hypothetical orbital shells to investigate more efficient potential satellite configurations. Additionally, each satellite can be further customized by defining the number of LISLs aboard. \par
Additionally, in order to simulate network functionality, SLASh contains a network communications simulator that can generate dummy data and simulate its movement throughout the network, allowing users to test their satellite shell configurations and compare their results with other SLASh simulations. In doing so, we successfully create an engine designed for public use that allows users to identify critical inefficiencies within LEO networks.


\subsection{Primer on Satellite Networks}

Every LEO satellite network contains a variation of inter-satellite link (ISL) configurations, technologies, and other variables necessary to define the topology and shape of each satellite network. This section serves as a brief but necessary introduction to the terminology used throughout this paper and further discussions on LEO networks. \par
The total collection of a satellite provider's satellite population in orbit is defined as a satellite constellation. Each satellite provider further divides their individual constellations into orbital shells, which consist of hundreds to thousands of satellites that all orbit Earth at a fixed and equivalent altitude. These shells rarely, if ever, communicate with each other, and primarily serve to allow numerous networks to exist while reducing the likelihood of collision or the requirement to make evasive maneuvers. As an example, the first generation of Starlink satellites were separated into five orbital shells, with each shell varying in altitude by roughly ten kilometers\cite{b8}.\par
Within each orbital shell, the satellites within are further divided into orbital planes. These planes typically divide the number of satellites equally among each plane, and akin to orbital shells, satellites do not typically change their orbital plane once they have been placed within a provider's constellation. These orbital planes are further defined by their angle of inclination, or the angle at which they orbit the Earth relative to the equator. In the event satellites orbit parallel to the equator, they have an inclination of 0\degree\cite{b9}.\par
Satellites can communicate with other satellites within the same orbital plane using laser inter-satellite links (LISLs), which serve a similar purpose to terrestrial internet infrastructure such as fiber-optic cables. LISLs allow satellites to rapidly transmit data through the vacuum of space, assuming the satellite receiving the data is within a direct line-of-sight. While there are other methods utilized by satellites for inter-satellite communications and data transfer, LISLs define the modern standard aboard many of the aforementioned satellite network providers, and throughout this paper we analyze satellite shells that utilize these links.\par
In order to communicate with terrestrial networks, satellites require a ground station to both receive and transmit data. These ground stations serve as a "gateway" to terrestrial Internet infrastructure and are necessary to connect with networks outside of the satellite shell, including the Internet. These ground stations typically consist of an array of dishes in order to access a wider range of satellites, allowing more than one satellite to connect to a ground station at a single time \cite{b16}.\par
\begin{figure}
    \centering
    \includegraphics[width=0.9\linewidth]{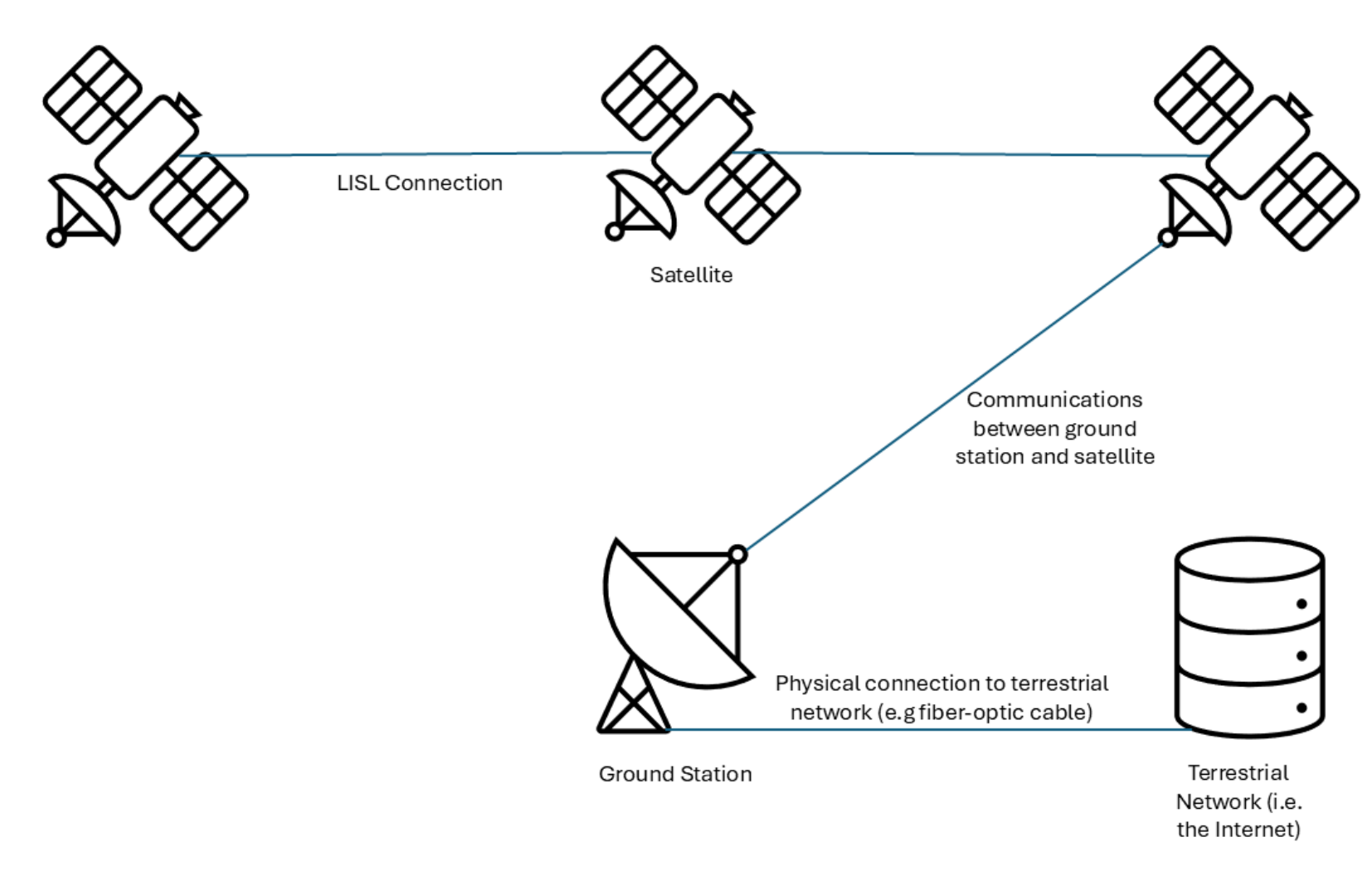}
    \caption{Diagram of a simple example network.}
    \label{fig:primer}
\end{figure}

Figure \ref{fig:primer} illustrates a simple example of what a LEO satellite network may look like. It is notable that the diagram shows a ground station with only one dish for simplicity, while in reality these stations are often equipped with multiple dishes to establish communications with more than one satellite in parallel.

 \section{Related Work}

As the importance of LEO Satellite Networks is growing, so has the amount of research being conducted across multiple institutions. To this end, several other LEO Satellite Network simulations have recently been developed. Some examples of such simulations include StarryNet \cite{b2} and Xeoverse \cite{b3}, two general-purpose LEO Satellite Network simulators. The term "general purpose" here describes simulations in which customization and adaptability are key features. Both StarryNet and Xeoverse were designed to be used by other researchers to make advancements in the LEO Network space without needing access to an existing real satellite network. This is extremely useful for researchers who are located in areas that don't have access to popular satellite networks (like StarLink) or for whom such access would be prohibitively expensive.\par
In addition to general purpose simulation platforms like StarryNet and Xeoverse, researchers have also recently developed more limited purpose simulations. The term "limited purpose" defines simulations that were built around a single research objective, such as studying a particular metric of a network. One example of a limited purpose simulation is the SaTE simulation \cite{b1}, which simulates an LEO Satellite Network for the express purpose of researching traffic engineering. Another example of a limited purpose simulation is the LeoEM \cite{b14} simulation, which was created as a testbed for SaTCP, a protocol made to improve congestion control in LEO Networks.

\subsection{StarryNet}

StarryNet \cite{b2}, while technically not a \textit{simulation} platform but rather an \textit{emulation} platform, is fundamentally a framework designed to simulate network-connected satellite constellations such as those deployed in StarLink, OneWeb, etc. It is written in Python and was developed by Tsinghua University in 2023. The primary motivation for this work was to design an emulation platform that was highly flexible and capable of realistic network simulation. This effort was specifically in an attempt to build an evaluation framework for future experiments on so-called "Integrated Space and Terrestrial Networks" or ISTNs \cite{b2}.\par
StarryNet uses containerization technology to represent different elements in the simulation (such as satellites and ground stations), and LISLs are represented as virtual network interfaces \cite{b2}. While this is scalable and performant, it adds additional complexity which can make it difficult for first-time users to understand.

\subsection{Xeoverse}

Xeoverse is another general purpose LEO Satellite Network simulation platform built for use by researchers \cite{b3}. It was written in 2024 by researchers at the University of Surrey. Unlike StarryNet, which is readily available on GitHub, xeoverse does not appear to have their codebase public, which may limit the amount of use it receives from other researchers. The primary motivation for Xeoverse was to improve on previous LEO Satellite Network simulations like StarryNet, and provide even more performance and scalability.\par
Instead of using containerization technology to represent network actors, xeoverse relies on Mininet, which is a process-level virtualization software \cite{b3}. This allows for emulating software-defined networks, such as those in LEO Satellite Networks.

\subsection{Other Simulations}

Other simulations of LEO Satellite Networks have been made to target specific research questions. The SaTE simulation \cite{b1} was developed by researchers at the University of Singapore and the University of Toronto specifically to study traffic engineering in LEO Satellite Networks. The simulation used for SaTE is not an experimentation testbed like StarryNet and Xeoverse, but rather a purpose-built solution that models 4 completed orbital shells of the StarLink network. In addition to LISLs, inter-satellite communication in the SaTE simulation is also able to leverage ground-relay stations (called "bent-pipe"). Satellites in the simulation contain a fixed number of 4 LISLs.\par
The LeoEM simulator \cite{b14}, like SaTE simulation, was designed for testing an application of particular research software. In this case, the simulation was built to test a congestion control algorithm called SaTCP. The project was written by researchers at the University of California San Diego in 2023, and is written in a mixture of Python, C, and MATLAB. Although LeoEM was built for a specific use-case, it is fairly robust and adaptable for a variety of uses. LeoEM also features the "bent-pipe" transimission capability seen in the SaTE simulation in addition to LISL capability.

\subsection{Motivation for SLASh}

Given the variety of previous LEO Satellite Network simulations available from recent research, the question is raised concerning the motivation for the development of SLASh. The primary goal of SLASh is to be a simulation platform with an express focus on LISL communication (e.g. satellite-to-satellite data transfer), rather than emulating the entire complex network system. While other simulation platforms mostly include LISL communication functionality, it is not the central focus. We propose SLASh, therefore, to be a simulation platform in which experimentation with LISL properties is configurable for the end user.\par
Further, the aforementioned simulation platforms all face a usability difficulty. Given the nature of research software, the execution and behavior of these simulation platforms, especially complex platforms like StarryNet, lend themselves to a steep learning curve for the end user. Given the narrowed scope of the SLASh project, it becomes achievable to have an approachable and user-friendly interface that would conceivably allow for increased adoption.

\subsection{Broader Research Context}
While SLASh focuses on the simulation of LISL communication, it is important to situate this work within the broader landscape of satellite network research. Recent advancements have expanded beyond routing and simulation to address critical challenges in security, computing infrastructure, and energy efficiency, all of which are vital parameters for future high-fidelity simulations.

For instance, the physical constraints of LEO satellites have driven research into optimizing energy efficiency for protocols like LoRaWAN \cite{shergill2024energyefficientlorawanleo} and managing thermal loads for on-board computing \cite{yuan2024heatsatellitesmeatgpus}, as hardware limitations directly impact network performance. From a security perspective, as satellite networks integrate with terrestrial systems, new vulnerabilities have emerged in areas ranging from 5G integration \cite{ali2023security5gnetworks} and ML-based secure communications \cite{song2022mlbasedsecurelowpowercommunication} to the security of microservices and OS-level virtualization utilized in software-defined satellites \cite{gopal2022securityprivacychallengesmicroservices, ketha2025analysissecurityoslevelvirtualization}.
Furthermore, the massive volume of telemetry data generated by these constellations requires robust processing architectures. Innovations in heterogeneous computing systems \cite{khatri2022heterogeneouscomputingsystems}, parallel data processing (e.g., MapReduce) \cite{qiu2023mapreducemultiprocessinglargedata}, and efficient data classification \cite{dixit2023dataclassificationmultiprocessing} are becoming essential for managing the next generation of satellite networks. The SLASh platform aims to provide a streamlined testbed that can eventually support the evaluation of these evolving technologies.

\section{Design}

For the implementation of the simulation, we considered a variety of languages and frameworks that would best suit our needs. We settled on using Python for our implementation because of its versatility, user-friendly nature, and rich library ecosystem. SLASh leverages three open-source libraries to assist with behavior involving orbital mechanics, network topology, and event simulation, which are detailed in the sections that follow.\par
The overall simulation flow is as follows:
\begin{enumerate}
    \item Generate a network based on satellite JSON files.
    \item Generate data flows each representing data traveling between two satellites.
    \item Simulate data moving through the network along the shortest-path until all flows have completed.
    \item Report statistics of the simulation.
\end{enumerate}

The simulation requires a directory of JSON files to be populated with information about the satellites. These are generated once via a generation Python script. Satellite parameters must be tuned in that script before the script is ran.

\subsection{Program Architecture}\label{AA}


\begin{figure}
    \centering
    \includegraphics[width=1\linewidth]{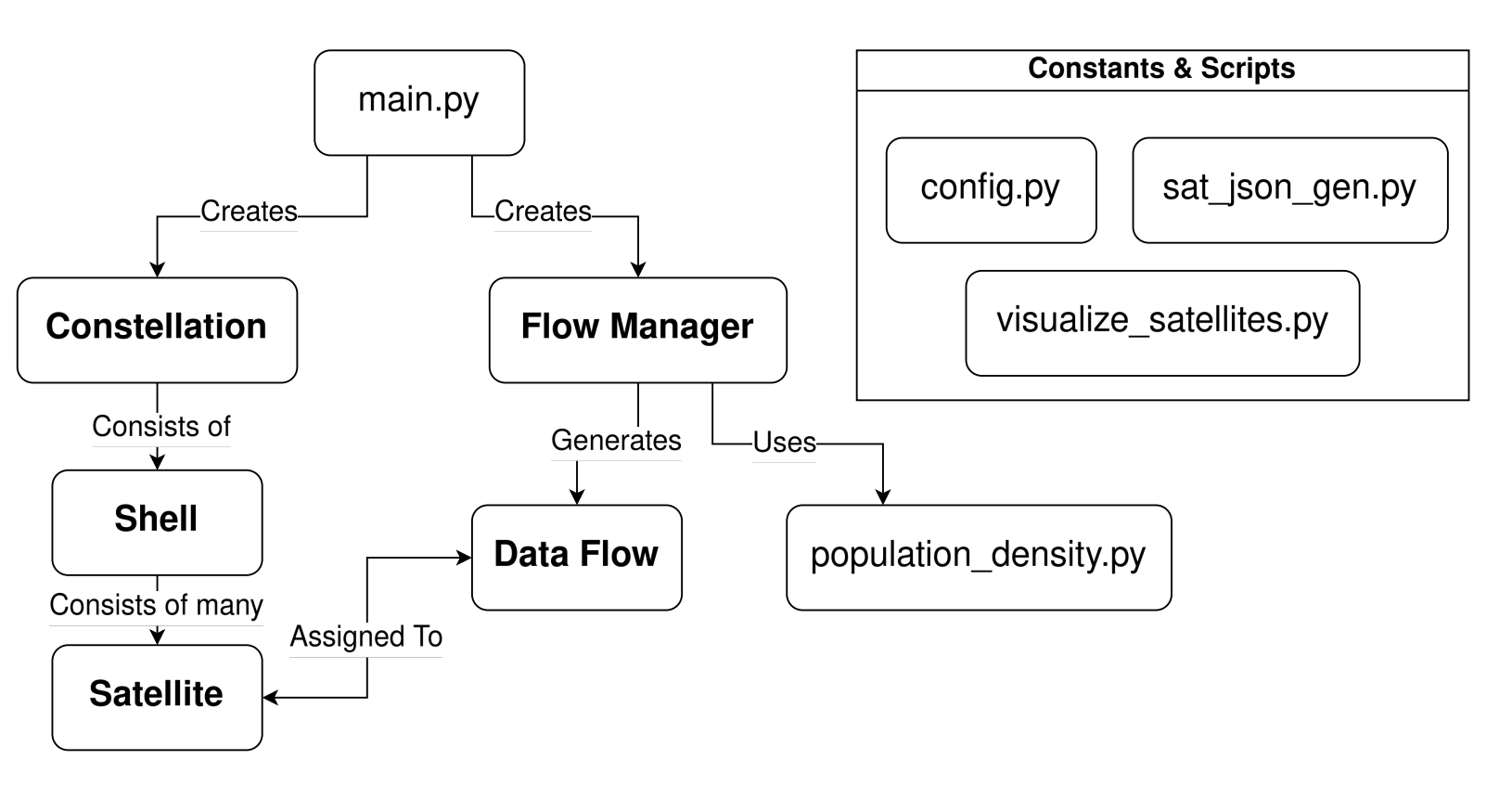}
    \caption{SLASh Program Architecture}
    \label{fig:Program Architecture}
\end{figure}

The simulation consists of a variety of Python class files and Python scripts. Figure \ref{fig:Program Architecture} shows a visual representation of the program architecture with class files in bold. The entrypoint to the program for the user is the main.py script, which generates the constellation from satellite JSON files and procedurally generates data flows to assign to the satellites. During execution, data flows are generated, assigned to satellites, and the simulation is carried out.\par
The sat\_json\_gen.py script is used to generate the JSON files for the satellites in the network. Each JSON file contains a numerical id for the satellite, and references to it's neighboring satellites that establish the topology of the network.\par
The config.py script contains definitions for important simulation parameters, like the number of orbital planes, number of satellites per plane, and constants used in the network simulation. This file becomes a key interface for tuning the simulation behavior, making the program user-friendly and customizable.\par
Finally, the visualize\_satellites.py script is used to generate visualizations for the satellite network, including an interactive visualization of satellites placed around a globe, as well as 2D graphs for satellite placement and satellite "priority", a concept explained in detail in section \ref{AE}.

\subsection{Orbital Mechanics}\label{AB}


The nature of LEO Satellite Networks necessitates that orbital mechanics play a large role in any attempt at simulation. This is because most of the unique characteristics of an LEO Network have to do with orbital motion and the placement of satellites around the globe. To assist in the calculations of orbital motion, SLASh leverages the Poliastro Python Library \cite{b4}. Poliastro is a capable and complex open-source library for Python that allows SLASh to define orbits for its satellites and place the satellite on the orbit. It also allows for calculations of distance and Earth-relative location (e.g. longitude and latitude) that are used to determine traffic levels for the satellite (making satellites in more populated areas have more traffic to handle).\par
SLASh defines a circular Earth-centered orbit for each individual satellite. The particular orbit is determined by the orbital plane in which the satellite sits. By default, SLASh generates 72 orbital planes, with 22 satellites in each plane. This number was chosen to mimic the topology of Shell 4 of the StarLink Network, since all satellites in that shell contain LISLs, unlike previous StarLink shells \cite{b13}. SLASh also places satellites with a degree of inclination of 53\degree\ by default, following the Shell 4 specification.

\subsection{Network Topology}\label{AC}

\begin{figure}
    \centering
    \includegraphics[width=0.75\linewidth]{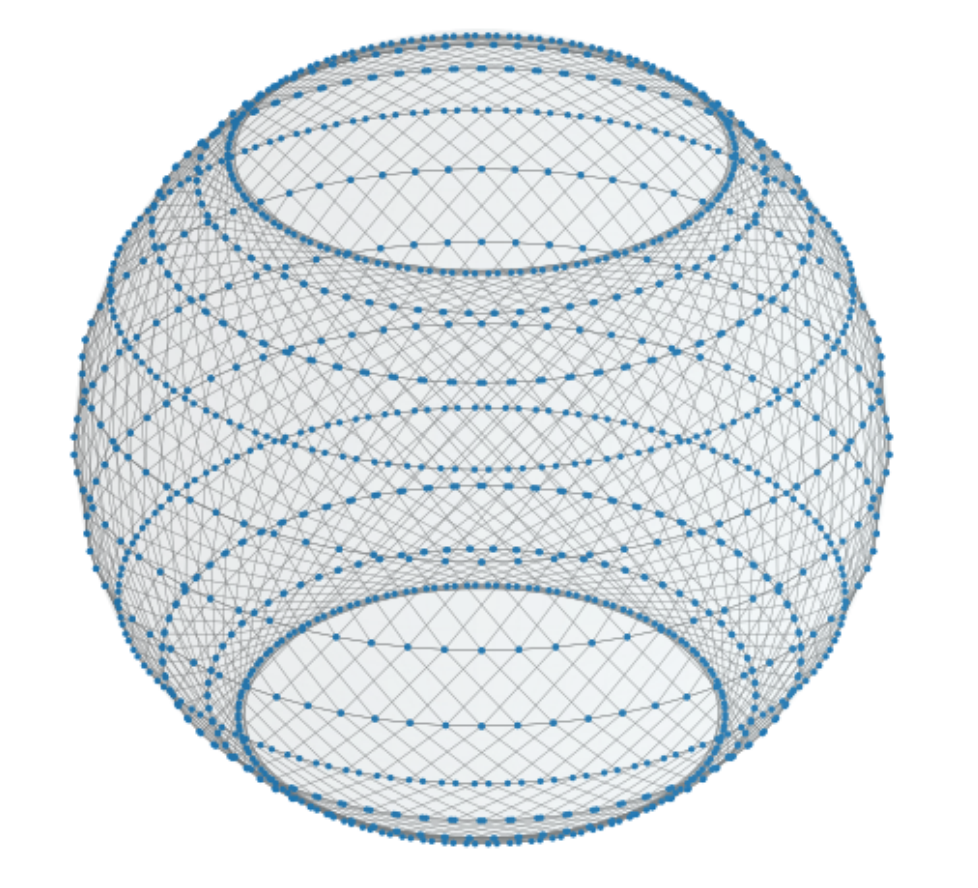}
    \caption{Visualization of Satellite Network Topology}
    \label{fig:Topology}
\end{figure}

To facilitate the simulation of network connectivity within the satellite shell, SLASh uses the NetworkX Python library. NetworkX allows users to create complex networks with ease due to its simple interface and collection of class methods. In addition, NetworkX is closely related to the SciPy library, simplifying the interactions between the different aspects of SLASh \cite{b17}.\par
When generating the JSON files, sat\_json\_gen.py assigns each satellite various identifying information including an 1-indexed ID, an altitude to determine which shell the satellite belongs to, the number of LISLs aboard, and a list of IDs assigned to neighboring satellites via the generate\_satellite\_files method. The final key-value pair is determined algorithmically allowing the user to create satellite constellations with thousands of nodes with ease. By default, the program assumes a set of four neighbors per satellite. This list of neighbors include the two adjacent satellites within the same orbital plane, as well as the two corresponding satellites in neighboring planes. \par
To determine the numeric IDs for the list of neighbors, we further determine a 0-indexed plane number \(p\) and a 0-indexed plane index \(i\). Additionally, the generate\_satellite\_files method is passed the number of orbital planes and the number of satellites within each orbital plane, which are represented as \(o\) and \(q\) respectively. We can then use the formula 
\[n_1=pq + ((i-1)\%q)+1\]
where \(n\textsubscript{1}\) is the ID of the previous\footnote{We use the terms "previous" and "next" here due to the decremented and incremented index of the satellite, and later, the orbital plane, and does not necessarily imply graph direction.} neighboring satellite within the same shell. Similarly, the other three neighbors can be calculated via the following formulae:
\[n_2=pq+((i+1)\%q)+1\]
\[n_3=q((p-1)\%o)+i+1\]
\[n_4=q((p+1)\%o)+i+1\]
These defaults mimic the specifications we designed when modeling our simulation after Starlink Shell 4. \par
In order to create the network topology within the program, we used a two-pass approach. The first pass consists of creating nodes for each of our satellite abstractions stored within the JSON files generated by sat\_json\_gen.py. Using NetworkX's flexible node system, where any hashable object may be used as a node in a generated graph \cite{b17}, we generate a Satellite object for each JSON file and add it to a graph based on the satellite's altitude attribute with the add\_node() method. Then, we take a second pass through each JSON file, establishing the edges within each shell by iterating through the list of neighbors and the add\_edge() method. This approach allows us to generate multiple shells at once, dependent on the altitudes present in the provided JSON files, while ensuring each shell is a separate and isolated graph. \par
Figure \ref{fig:Topology} shows a visualization of the topology of the network. Each blue vertex represents one satellite (or node) in the network, while each gray line represents the LISL connection between two satellites (the edge).

\subsection{Traffic Generation}\label{AE}

\begin{table}
    \centering
    \begin{tabular}{cc}
        \textbf{Parameter} & \textbf{Method of Generation} \\
        \hline \\
         Identifier & Sequential \\
         Source Satellite & Selected based on Population \\
         Destination Satellite & Selected based on Population \\
         Size In Bytes & Randomly Chosen from a Range \\\\
    \end{tabular}
    \caption{Data Flow Parameters and Their Generation}
    \label{tab:Data Flow Parameters}
\end{table}

\begin{table}
    \centering
    \begin{tabular}{cccc}
        \textbf{Region} & \textbf{Latitude Range} & \textbf{Longitude Range} & \textbf{Weight} \\
        \hline \\
        North America & [25, -50] & [-130, -60] & 8.0 \\
        Europe & [35, 60] & [-15, 45] & 9.0 \\
        Asia & [10, 55] & [60, 150] & 9.5 \\
        Middle East & [12, 38] & [30, 65] & 5 \\
        South America & [-55, 15] & [-85, -30] & 4.0 \\
        Africa & [-35, 37] & [-20, 55] & 3.0 \\
        Oceania & [-50, -10] & [110, 180] & 2.0 \\\\
    \end{tabular}
    \caption{Traffic Regions and Weight}
    \label{tab:Traffic Regions}
\end{table}

\begin{figure}
    \centering
    \includegraphics[width=1\linewidth]{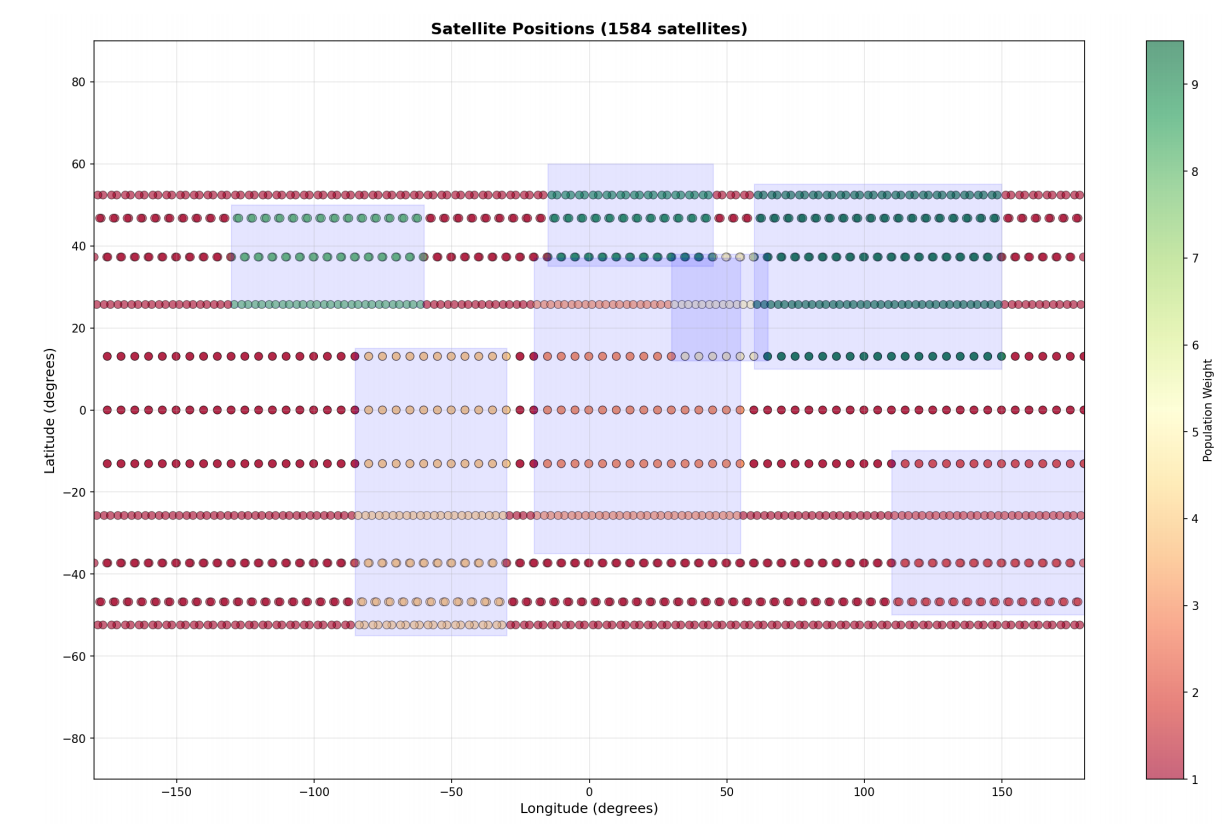}
    \caption{2D Plot of Traffic Regions}
    \label{fig:Traffic Regions Map}
\end{figure}


Once the network topology is established, SLASh begins generating data flows to represent traffic traveling throughout the network. Data flows are generated all at once before the simulation is begun. Table \ref{tab:Data Flow Parameters} shows the key parameters for each data flow. Namely, data flows contain a unique identifier, as well as a source and destination satellite, and a size in bytes. The identifier is generated sequentially so each data flow remains unique. The source and destination satellites are selected through a weighted distribution based on region.\par
Table \ref{tab:Traffic Regions} shows the definition of traffic regions and their respective weights. These are rectangular regions based roughly around the position of various continents and populations around the globe. Fluctuations in traffic demand throughout the regions are not considered. The simulation therefore has a rudimentary population-based preference for traffic, but is by no means comprehensive. This approach to traffic generation was decided on as a tradeoff between a realistic traffic generation system and the time constraints of the project.

\begin{figure}
    \centering
    \includegraphics[width=1\linewidth]{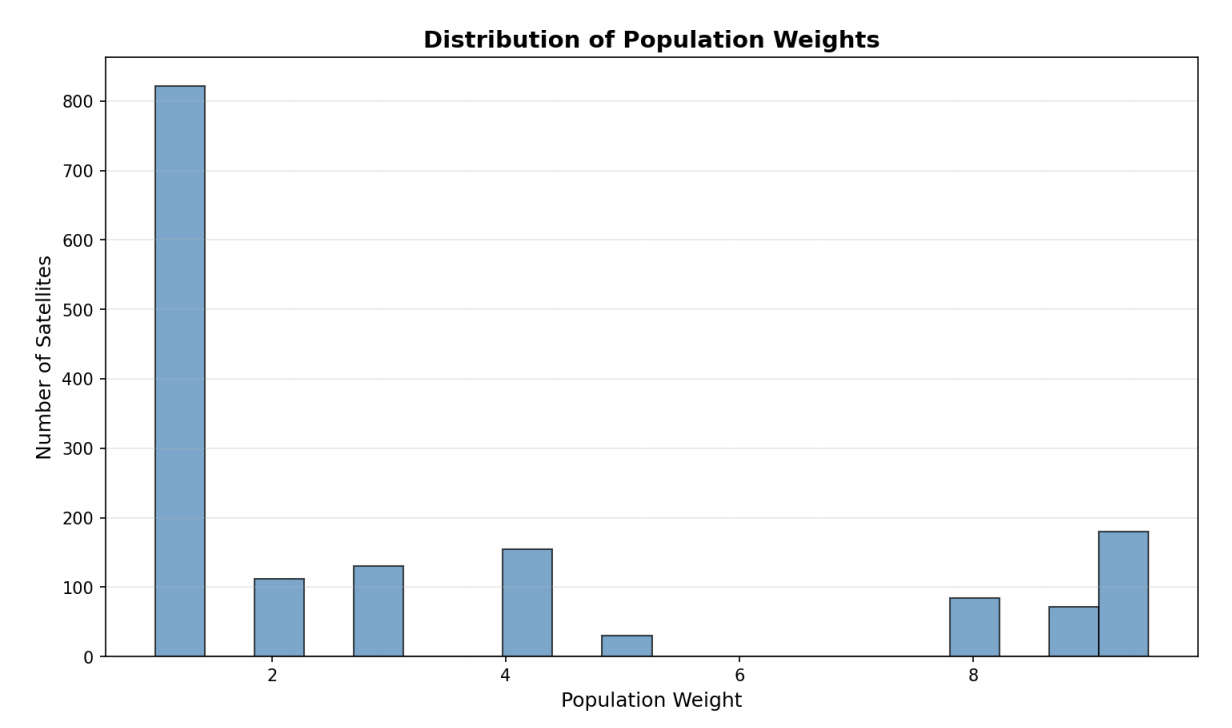}
    \caption{Weight Distribution of Satellites}
    \label{fig:Weight Distribution}
\end{figure}

Figure \ref{fig:Traffic Regions Map} shows the traffic generation regions plotted on a graph of longitude and latitude, as well as the satellites in the network. Satellites are color-coded to highlight their weight between 1 and 9. Satellites that are weighted higher are chosen more often as either source or destination satellites, mimicking real-world behavior as global networks contain more traffic where there is a higher population. The weights themselves were chosen to roughly match population demographics and internet usage per region \cite{b5, b6}.\par
Figure \ref{fig:Weight Distribution} shows the distribution of satellites across weight of their region. It can be seen that a vast majority of satellites have a low-weight, which follows given the satellites' distribution across the globe and the relative sparsity of population demographics. Most of the satellites are positioned on open water, or in lowly populated areas, and so the distribution highly favors low weighted satellites. The primary goal for the region-based traffic generation system is to try and devalue these satellites to create a more realistic network environment.\par
The number of data flows that are generated varies between execution. In general, the user inputs an amount of data they would like to simulate (e.g. 1MB, 100GB, 3TB, etc.) as well as a range for sizes of data flows (by default, this range is between 1KB and 10MB). SLASh then repeatedly generates data flows with a size uniformly selected from the range until it reaches the total data amount. This, along with the weighted selection of source and destination satellites, provides all the necessary information about data flows.

\begin{figure}
    \centering
    \includegraphics[width=1\linewidth]{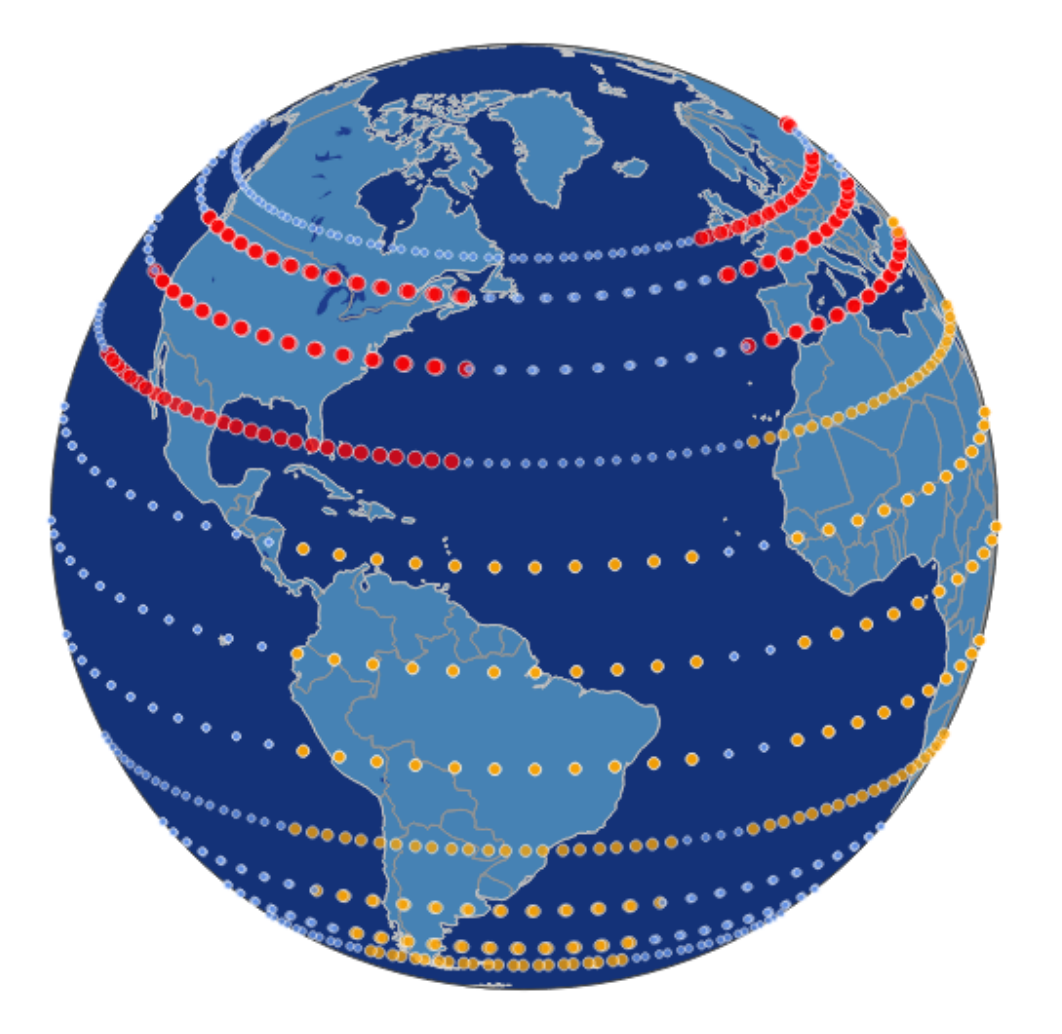}
    \caption{Network Topology with Weighted Satellites around a Globe}
    \label{fig:Topology with Weights}
\end{figure}

Figure \ref{fig:Topology with Weights} shows a visualization of the generated network topology with satellites colored based on their weight. It should be noted that, as the traffic regions are rectangular, they do not perfectly align with continental boundaries or coastlines. Since the weight of an individual satellite represents it's likelihood of selection during data flow generation, and not the total amount of data traveling through that satellite, this level of inaccuracy to continental boundaries is acceptable.

\subsection{Network Event Simulation}\label{AD}


After a network topology has been generated and data flows have been constructed, actual simulated events begin happening. Simulation behavior in SLASh leverages the SimPy Python library \cite{b15}, which is a discrete event simulation platform. In essence, SimPy allows for the creation of simulated "discrete events" through use of standard Python generator functions. These events (or \textit{processes} as they are called in SimPy terminology) can then be scheduled and replayed in simulation-time. This allows for a relatively straightforward simulation paradigm in which discrete network events are created and scheduled to represent data being transferred across satellites.\par
The primary SimPy process that handles data transference between satellites is called the "scheduler" process. This process enters a loop in which satellites wait for incoming data, buffer data that comes in in chunks, and then passes that data on to the next hop. Each satellite is capable of sending data to any of its neighboring nodes as defined by the network topology, and each LISL sends at a bitrate defined in the user's configuration settings (by default this is 100Gbps, the maximum bitrate of StarLink Shell 4 LISLs).\par
As flows reach their destination satellite, the satellite reports the amount of data that has reached its destination. When the flow is completely finished, the satellite reports to the simulation process that the flow is completed, and a log line is outputted to the terminal output. A progress bar is also shown during the simulation which signals to the user how much time is remaining in the simulation as well as how much progress has occurred thus far. When all data flows have been completed, the simulation ends. The simulation then reports the number of data flows, how many flows were dropped (if any), the amount of data transferred, the simulation time, and the real time that the simulation took to execute.

\subsection{Routing}

As flows travel from satellite to satellite in an attempt to reach their destination, each individual satellite needs to know where to route the data that is coming in. SLASh handles this by using a centralized routing system handled by the shell class. This is analogous to a centralized terrestrial routing system, similar to the one discussed in \cite{b1}. This routing system relies on the NetworkX topology and calculates the shortest path between the source and destination satellites. These routes are then cached, to avoid recalculation as the flow travels from satellite to satellite. This results in an efficient and performant system for routing data flows.

\section{Evaluation}

To evaluate the performance of a generated network, the simulation generates the statistics mentioned at the end of section 3E. We can then further compare the output from a variety of LISL arrangements and satellite shell topologies. This section enumerates the experimental process and the results of a basic SLASh simulation comparing possible LISL arrangements aboard a satellite shell that emulates Starlink's Shell Four.

\subsection{Experimental Setup}

Example simulations were run on a Microsoft Azure virtual machine, utilizing free credits obtained through the Azure for Students program. The virtual machine was set up to run Linux Ubuntu 22.04 on x64 architecture, using primarily default settings. The core uses the Standard D2s v3 package, with 2 vCPUs and 8GB of RAM, and is connected to a 1 TB data disk in addition to its OS disk. \par
Connection to the aforementioned virtual machine was performed on the personal computer of one of the authors, a Dell Precision 3430. Connection was established via SSH protocol, where the authors could remotely operate the virtual machine. \par
All testing was performed in a virtual Python environment on the virtual machine. This environment was set up on Python version 3.11, and utilized Python library versions provided with the source code. \par
To ensure testing would compare the efficiency of LEO satellite networks solely based on the adjustment of LISLs aboard each satellite, we created four arrangements of LISLs that were applied across all satellites in each given testing group. The first group acted as a control group, and was based on our provided knowledge of each satellite housing four LISLs on board. Additionally, we created two groups with three LISLs on board each satellite. For these groups, we removed the "previous" neighbor from the same shell and the adjacent shells, respectively. Finally, we tested the efficiency of the network when each satellite is equipped with only two LISLs, allowing for single-direction movement along each orbital plane as well as between orbital planes. Each group was tested over three trials to account for potential derivation in simulation time resulting from the generated telemetry data. \par
Regarding the generated dummy data, each simulation attempted to transmit 1 GiB (1024 MiB) throughout the virtual network. This data was broken up into some number of data flows as outlined in section 3D, utilizing system defaults to create flows between 1KB and 10MB, creating an estimated \~200 flows for each test. The number of data flows per trial, simulation time, and real time were recorded and are presented in the following subsection.

\subsection{Results}

The results of the simulation experiments are shown in Tables \ref{tab:Data 4LISLs}, \ref{tab:Data 3LISLs Within}, \ref{tab:Data 3LISLs Between}, and \ref{tab:Data 2LISLs}. Each table shows the statistics for each trial of the run, with the primary focus being on simulation time and real time for data to be processed. Each table highlights a different LISL arrangement onboard each satellite, which highlights SLASh's capability for LISL configuration.

\subsubsection{Arrangement 1 (4 LISLs)}

\begin{table}
    \centering
    \begin{tabular}{cccc}
         & Number of Data Flows & Simulation Time & Real Time\\
         Trial 1 & 217 & 0.82s & 17.67s\\
         Trial 2 & 224 & 0.81s & 18.93s\\
         Trial 3 & 207 & 0.83s & 18.66s\\
         \textbf{Average} & \textbf{216} & \textbf{0.82s} & \textbf{18.42s}\\
    \end{tabular}
    \caption{Data for LISL Arrangement 1 (4 LISLs)}
    \label{tab:Data 4LISLs}
\end{table}

\begin{figure}
    \centering
    \includegraphics[width=0.75\linewidth]{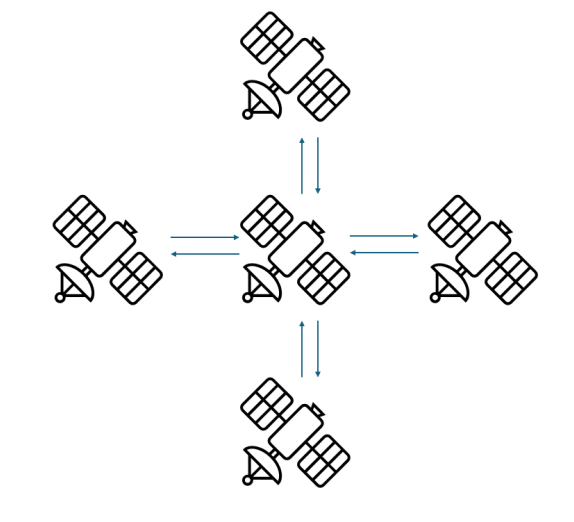}
    \caption{Visualization of Neighbors for Arrangement 1}
    \label{fig:Arrangement 1}
\end{figure}

For the initial execution of the experiment, a network topology was constructed with 4 LISLs onboard each satellite. As we modeled this execution after StarLink Shell Four, we simulate 1,584 satellites organized in 72 orbital planes with 22 satellites in each plane. This aligned with LISL configurations found in other research \cite{b1}, although this number may be inaccurate to real-world implementations like StarLink \cite{b13}. The results of the execution of this simulation run can be seen in Table \ref{tab:Data 4LISLs}. Each row shows an individual trial of the simulation, and the final row shows an average of the results from each trial.\par
As shown in the table, the simulated time is significantly smaller than the real time that the execution took. This is because of the SLASh's discrete event simulation behavior through SimPy, which allows the satellite scheduler process to run in a simulated time frame. Each run of the simulation has a different number of data flows, due to the behavior discussed in \ref{AE}, wherein the fluctuations of the sizes of individual data flows contributes to a different number of final data flows as the amount of data being process is fixed across trials.\par
The average simulation time for the entire simulation to finish is under a second, which is largely due to the configured transmission rate of individual LISLs. The real time taken is much larger, likely due to overhead from producing simulated events, calculating routing decisions, and buffering data. \par
A basic visualization of the arrangement of satellites and the direction in which data can be transmitted can be found in Figure \ref{fig:Arrangement 1}. Additionally, the maximum number of steps a data flow may take can be calculated by the formula \(s=(o+q)/2\).

\subsubsection{Arrangement 2 (3 LISLs, 1-Directional Within Plane)}

\begin{table}
    \centering
    \begin{tabular}{cccc}
         & Number of Data Flows & Simulation Time & Real Time\\
         Trial 1 & 216 & 0.81s & 16.76s\\
         Trial 2 & 210 & 0.84s & 17.19s\\
         Trial 3 & 197 & 0.81s & 17.09s\\
         \textbf{Average} & \textbf{207.67} & \textbf{0.823s} & \textbf{17.01s}\\
    \end{tabular}
    \caption{Data for LISL Arrangement 2 (3 LISLs, 1-Directional Within Plane)}
    \label{tab:Data 3LISLs Within}
\end{table}

\begin{figure}
    \centering
    \includegraphics[width=0.75\linewidth]{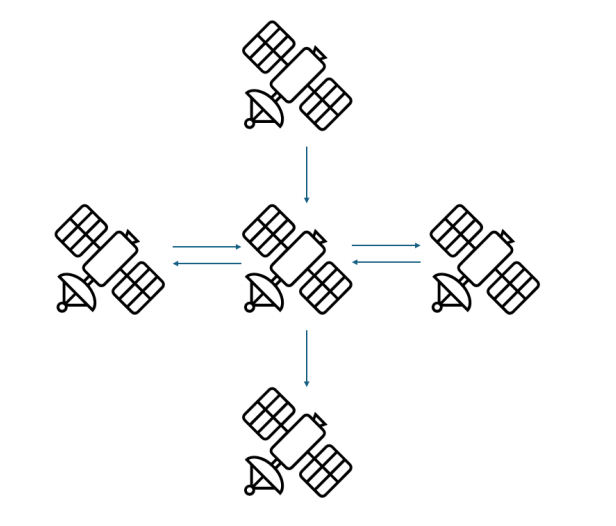}
    \caption{Visualization of Neighbors for Arrangement 2}
    \label{fig:Arrangement 2}
\end{figure}

In this arrangement, we generated a the same amount of satellites as before, also organized in 72 orbital planes with 22 satellites within each plane, again modeled after Starlink's Shell Four. Unlike the last arrangement, one LISL has been removed from each satellite, which was accomplished by removing the neighbor ID assigned to the previous neighbor. This representation may more accurately represent the true alignment of LISLs aboard Starlink's Shell Four. The data for the tests resulting from this arrangement can be found in Table \ref{tab:Data 3LISLs Within}.\par
As shown in the table, the real time execution for running the simulation is over a second lower than it was with arrangement 1. This is likely due to less simulation overhead as data is passing between satellites, because fewer LISLs means fewer possibilities for routes and lower amounts of processing per individual satellite. However, the simulation time is relatively constant, demonstrated similar performance for the network with fewer LISLs. \par
A basic visualization of the arrangement of satellites and the direction in which data can be transmitted can be found in Figure \ref{fig:Arrangement 2}. While the illustration does not visualize the entire network, it is important to note that the circular nature of each orbital plane allows data to transfer from any one satellite within a plane to another, despite the directional limitation, by traversing the entire plane until the destination satellite is reached. This means the maximum number of steps a data flow must take can be calculated by \(s=o/2+q-1\).

\subsubsection{Arrangement 3 (3 LISLs, 1-Directional Between Planes)}

\begin{table}
    \centering
    \begin{tabular}{cccc}
         & Number of Data Flows & Simulation Time & Real Time\\
         Trial 1 & 186 & 0.82s & 16.91s\\
         Trial 2 & 216 & 0.82s & 16.68s\\
         Trial 3 & 223 & 0.83s & 16.98s\\
         \textbf{Average} & \textbf{208.33} & \textbf{0.823s} & \textbf{16.88s}\\
    \end{tabular}
    \caption{Data for LISL Arrangement 3 (3 LISLs, 1-Directional Between Planes)}
    \label{tab:Data 3LISLs Between}
\end{table}

\begin{figure}
    \centering
    \includegraphics[width=0.75\linewidth]{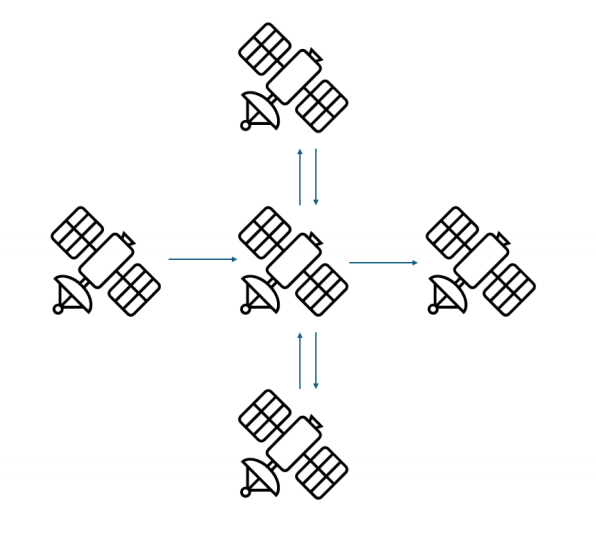}
    \caption{Visualization of Neighbors for Arrangement 3}
    \label{fig:Arrangement 3}
\end{figure}

Similarly to the previous arrangement, this satellite network is organized with the above listed parameters for satellite placement. However, unlike arrangements 1 and 2, arrangement 3 lacks the ability to traverse in both directions between orbital planes. This was accomplished by removing the neighbor ID assigned to the corresponding satellite in the previous plane. Akin to arrangement 2, this arrangement may more accurately represent the true alignment of LISLs aboard Starlink's Shell Four. The data for the tests resulting from this arrangement can be found in Table \ref{tab:Data 3LISLs Between}. \par
A basic visualization of the arrangement of satellites and the direction in which data can be transmitted can be found in Figure \ref{fig:Arrangement 2}. Akin to Arrangement 2, the circular nature of the orbital planes in reference to one another allows data to transfer from any satellite to its destination. The maximum number of steps in this case can be calculated using the formula \(s=o-1 + q/2\). It is notable that in our example, this value is greater than the maximum number of steps possible for Arrangement 2's shortest distance, as \(o>q\).

\subsubsection{Arrangement 4 (2 LISLs, 1-Directional Between and Within Planes)}

\begin{table}
    \centering
    \begin{tabular}{cccc}
         & Number of Data Flows & Simulation Time & Real Time\\
         Trial 1 & 209 & 0.80s & 16.88s\\
         Trial 2 & 216 & 0.82s & 16.08s\\
         Trial 3 & 225 & 0.83s & 17.15s\\
         \textbf{Average} & \textbf{216.67} & \textbf{0.817s} & \textbf{16.70s}\\
    \end{tabular}
    \caption{Data for LISL Arrangement 4 (2 LISLs, 1-Directional Between and Within Planes)}
    \label{tab:Data 2LISLs}
\end{table}

\begin{figure}
    \centering
    \includegraphics[width=0.75\linewidth]{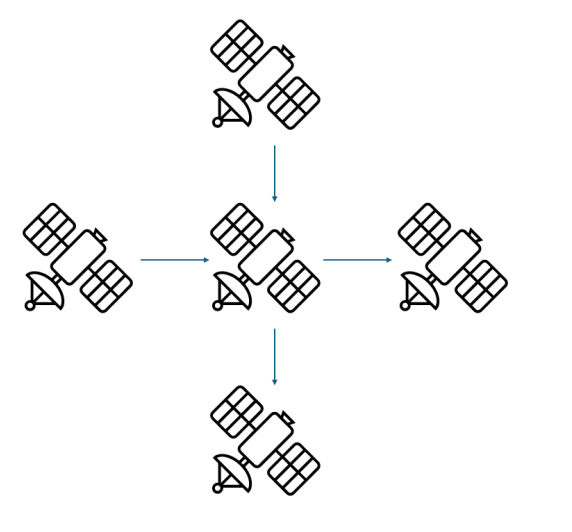}
    \caption{Visualization of Neighbors for Arrangement 4}
    \label{fig:Arrangement 4}
\end{figure}

Finally, we tested a hypothetical arrangement of LISLs where each satellite in the network only contains two LISLs. This arrangement was selected due to a natural progression of the decrease in LISLs aboard each satellite. Like arrangements 2 and 3, was created by removing both the previous neighbor within the same plane and the corresponding satellite in the previous plane. The data for the tests resulting from this arrangement can be found in Table \ref{tab:Data 2LISLs}, and a visualization of the neighbors of any satellite can be found in Figure \ref{fig:Arrangement 4}. The maximum number of steps a data flow must take in this case is \(s=o+q-2\).

\subsubsection{Overall Results and Interpretation}

\begin{table}
    \centering
    \begin{tabular}{cccc}
         & Number of Data Flows & Simulation Time & Real Time\\
         Arr. 1 & 216 & 0.82s & 18.42s\\
         Arr. 2 & 207.67 & 0.823s & 17.01s\\
         Arr. 3 & 208.33 & 0.823s & 16.88s\\
         Arr. 4 & 216.67 & 0.817s & 16.70s\\
    \end{tabular}
    \caption{Average Data for Four LISL Arrangements}
    \label{tab:Data Averages}
\end{table}

A comparison of the average of all four LISL arrangements can be found in Table \ref{tab:Data Averages}. As evident within the table, the arrangement of LISLs aboard each satellite has effectively zero effect on the network's ability to transmit data when the amount of LISLs is greater than or equal to two.\par 
The arrangements with fewer LISLs aboard each satellite took over one second less time to complete the simulations in real time. We theorize this is likely due to the routing cache mentioned in Section III-F. Since there are fewer possible routes for any given data flow to take, it is logical that any pathing is more likely to generate a hit from a more restricted graph than a less restricted one. This theory is corroborated by the acceleration of the simulation as it runs, which is not represented in the data. As the simulation runs, the rate at which flows are completed accelerates greatly, which can be seen in both the progress bar and the provided rate calculator visible when running the simulation. This effect is more dramatic when there are more flows, further providing evidence that the routing cache has a significant effect on the real time duration each simulation has, and thus is not relevant to the LISL count aboard each satellite. \par
The throughput of the satellite network seems to be less related to the number of steps each data flow had to take and is more closely related to other factors, such as the overall bitrate of any given LISL. It is trivial that adding more powerful transmitters to a given network would increase the throughput of the network. Due to the abstract representation this simulation relies on, more specific knowledge on the exact specifications of LISLs will allow researchers using SLASh to return varied results.

\section{Issues}

Throughout the development process of SLASh, we encountered hurdles and challenges that slowed down overall development and demanded costly research sessions. These hurdles often came in the form of a lack of familiarity with the subject matter, due to our relative inexperience with topics such as orbital physics or realistic insights into network traffic behavior. These issues are described below.

\subsection{Reliable Information Regarding Satellite Constellations}\label{BA}

Due to the lack of transparency and absence of public consensus on the exact specifications of certain satellites, we were left no other choice than to make assumptions regarding the structure of each satellite. For example, in our testing regarding satellites present in Starlink's Shell 4, we defined each satellite as having four LISLs as our research returned that specification \cite{b12}. However, as parallel research performed by Dr. Zhu's lab uncovered, the Starlink website defines each satellite as having three LISLs on-board \cite{b13}, which contradicts our prior research with data from a more reliable source. While this does not affect the overall usefulness or accuracy of SLASh itself, it does highlight the lack of transparency LEO constellations have in the public eye, and furthermore raises questions on the accuracy of public data pertaining to satellite constellations as a whole. \par
Additionally, due to the above limitation regarding public information on these satellites, we were unable to determine the actual neighbor-determining algorithms present aboard real-world LEO satellite shells. To circumvent this issue, we made the assumptions listed above when determining which satellites would be assigned as neighbors to other satellites, which in turn may or may not lower the accuracy of our testing. We have decided that, in lieu of real-world specifications on the neighbor-assigning algorithms, it was vital to resort to our own assumptions to assign neighbors and create a functioning network topology. Should more accurate information surface on the assignment of satellite neighbors surface or be brought to the authors, SLASh can be altered to account for this oversight.

\subsection{Realistic Internet Traffic Generation}

Understanding and synthesizing global internet traffic patterns and behavior is a complex field of study. Despite the strong application to global satellite networks (and hence the importance to SLASh), the complexity of the topic and the constraints of the project led to compromises being made in the implementation in SLASh. Gathering reliable, real-world statistics of global internet utilization is difficult, and the authors' relative inexperience with the topic led to it being considered largely out of scope, since understanding and building this data into the simulation would have required significantly more time. \par
SLASh therefore assumes a rudimentary traffic generation system, in which rough geographical regions designate which satellites are more likely to be involved in data flow transfers. SLASh also assumes only fixed-size data transfers, which ignores more complex network operations, such as data streaming, which is becoming more prevalent on global networks. \par
Additionally, a high-level "flow-based" design was chosen for SLASh that abstracted many fine-grained network details into an easier to understand paradigm. While this aligned with SLASh's goal of being user-friendly and approachable, it does alight over some information that a packet-level simulation would be able to provide, although such a simulation would be more processing-intensive and less user-friendly.

\subsection{Understanding Orbital Motion and Mechanics}

As students primarily concerned with computer science, neither author had an intensive background in understanding astrophysics or the mechanics present within satellite constellations. As such, both authors had to resort to surface level understandings of issues including angles of inclination, orbital speed, and other physical factors present within the construction of satellite shells. We attest that the research performed on these topics of mechanics is accurate and suitable for the creation of SLASh, albeit a better understanding of these topics would benefit further development of SLASh, and allow us to more easily develop desired features detailed in the next section. \par
As outlined earlier in this paper, many of the orbital mechanics and overall physics including satellite placement are a result of utilizing the Poliastro Python library. This library was also new to both authors, and thus we primarily focused on the more elementary features of Poliastro to avoid delving too deep into orbital physics.

\subsection{Simulation Development Characteristics}

SLASh is a product of two authors with no prior experience developing simulation platforms. Because of this relative inexperience development proved difficult, and often slower than anticipated due to the unfamiliarity with the subject matter. All three primary libraries that SLASh takes advantage of (poliastro, NetworkX, and SimPy) were new to the authors, and understanding their functionality as well as fitting their behavior together into a cohesive product became the primary challenge. \par
In light of this issue, SLASh was developed incrementally following a minimum-viable-product model in which the simplest form of a network simulation was built, and features were slowly added onto it to improve the product and make the simulation more applicable to real-world analysis. More experience with simulation technology would likely have improved the speed of development, allowing the project to reach a larger scale before the end of the project was reached.

\section{Conclusion}

LEO Satellite Networks are becoming increasingly important in the world of wireless networking. As global connectivity increases, the usage of these networks will only increase, highlighting a vital need for understanding and experimentation with the technology. In particular, the orbital mechanics and unique topology of such networks in addition to the growing need for research call for the existence of simulation software that is approachable and adaptable for performing future research. To this end, we present SLASh, an approachable LEO Satellite Network simulation program with particular attention on inter-satellite communication. \par
The authors of this paper are proud to report the developmental progress of SLASh thus far. The project exposed both of us to new technologies and new insights with regards to wireless networks. We also recognize that further developments could be made to improve the final product, and hope that our work has helped for further the body of knowledge and resources surrounding this burgeoning technology. The following sections enumerate some of the knowledge that we gained throughout the project, as well as ideas for future development of the software.

\subsection{What Was Learned}

Completion of this project exposed both of the authors to new material that contributed to our education and broadened our horizons, providing us with great value. Some topics that were unfamiliar to us at the start which we were able to gain experience with, or skills that we were able to improve as part of the development process are as follows:

\subsubsection{LEO Satellite Mechanics}

Before beginning the project, both authors had minimal exposure to the realities of LEO Satellite Networks, though we had some understanding through reading the papers assigned in the CSE 5432 course. While these readings provided a foundational understanding, significant research remained for us to understand the topology and behavior of LEO Satellite Networks enough to be able to write a functioning network simulation. While there were challenges and barriers to understanding these mechanics (discussed in Section \ref{BA}), we were able to persevere and gain a deeper understanding about how LEO Satellite Networks truly function. This information is useful not only in the context of a wireless networking course but also in our personal lives as the technology becomes more commonplace.

\subsubsection{Discrete Event Simulation in Python}

Another topic that was wholly unfamiliar to the authors going into the project was how event simulation worked in Python. This proved to be one of the more difficult aspects of SLASh to understand and program, given the lack of understanding we had at the start. Discovering and researching the SimPy Python library, learning about Python generator functions, and determining what network event behavior were all activities that left us with more of an understanding about simulations. Debugging the code when things went wrong also helped solidify the concepts. This experience gave us more of an understanding of the Python language and how simulations work, which is knowledge that may be useful in further educational and professional settings.

\subsubsection{Collaborative Development}

An integral component of this project was the collaborative development between both authors. Equitable distribution of research, programming, and writing tasks can be difficult to achieve in practice. Although collaborative projects frequently suffer from miscommunications and mishaps, the experience gained from working together throughout this project helped us to improve our collaborative development skills, which will be vital for future projects. In particular, we gained experience delegating tasks and communicating our results, allowing us to work asynchronously while still remaining collaborative.

\subsubsection{Academic Research}

Finally, a fourth learning outcome of this project for the authors was further experience in how to conduct and perform academic research. Research is unlike a majority of coursework in an undergraduate curriculum, involving much more freedom and self-paced activity than traditional classes. As one of the authors is an undergraduate student, this project served as an enlightening introduction to the world of academic research, and the writing of this paper served as a departure from writing styles of previous coursework. Whether or not the authors continue in academic research, this project has taught us more about how computer science research is conducted.

\subsection{Future Work}

In recent years, the landscape of network research has expanded rapidly, driven by advancements in Machine Learning \cite{10.1145/3460120.3484766,9894326,9709070,9444204,ning2021benchmarkingmachinelearningfast,8832180,wei2025,8556807,8422243,chandrasekaran2022computervisionbasedparking,iqbal2021machinelearningartificialintelligence,pan2020endogenous}, Internet of Things (IoT) \cite{10.1145/3769102.3770620,290983,wire1,10017581, 9523755,9340574,10.1145/3387514.3405861,9141221,9120764,10.1145/3356250.3360046,8737525,8694952,10.1145/3274783.3274846,10.1145/3210240.3210346,8486349,8117550,8057109,10189210}, and security. Although this paper focuses on LEO satellite network simulation, it is important to acknowledge that future satellite networks will inevitably integrate with these emerging technologies. While SLASh proves to already be a helpful tool in investigating the efficiency of LISLs aboard LEO satellites, there are areas for improvement throughout that we were unable to investigate further due to the time and personnel restraints on the project. This section further explores potential improvements that SLASh would benefit from.

\subsubsection{Dynamic and Improved Network Topology}

As SLASh stands, it currently assumes a static network topology with each satellite having constant and consistent connection with each of its neighbors. While this abstraction represents an ideal scenario for hypothetical purposes, it lacks the nuance necessary for understanding the real-world complications that can arise with satellites in low Earth orbit. For instance, the simulation lacks the ability to introduce potential interruptions between satellites or even the disabling of a satellite from the network entirely. In order to be a noteworthy and useful simulation for scientific use, the program requires some implementation of these complications into the system. \par
One way these complications could be implemented would be through a random removal of certain satellite nodes and edges from the simulated network. This could be done by adding a variable to config.py that controls the chance of any given node or edge's removal from the final satellite shell graph. This implementation would assume that a satellite or it's connection's removal from the network is constant throughout the entire simulation, with the node or edge not being present for its entirety. \par
Additionally, further understanding of the neighbor selection process in real-world satellite constellations could highlight issues within the assumptions made in our neighbor selection algorithm. The current system assumes that neighbors (and thus, edges) do not change once assigned, however, we lack the knowledge of how a dynamic neighbor selection process may affect transfer rates throughout the network. If, for instance, one neighbor is disabled for the duration of the simulation, a satellite outfitted with a retargeting system may be able to utilize the no longer useful LISL by selecting a new neighbor. \par
Furthermore, within the simulation, each satellite remains in a fixed position in respect to the Earth's surface, effectively setting its orbital speed to 0 km/hr. This is inaccurate to the real world speed that LEO satellites must travel at in order to maintain a constant altitude. For example, Starlink satellites travel at over 27,000 kilometers per hour (7.5 km/s) \cite{b18}. While this is not as relevant for smaller simulations with a lower data size and/or fewer satellites, larger simulations would necessitate this consideration to be accurate to real-world conditions. Simulating this motion would allow SLASh to create more realistic projections regarding the routing that the data flows take, as the satellites connected to each ground station change dynamically in reality.\par
In a separate vein from a dynamic topology, there are still improvements that can be made to the static topology generator. For instance, there is no system in SLASh that checks whether a satellite shell is physically possible due to line-of-sight restrictions. For instance, a hypothetical satellite constellation consisting of only one shell of two satellites should be rejected as there is no possible way to evenly space out those satellites while retaining line-of-sight due to the opaqueness of the Earth. This would require a more complex understanding of orbital geometry and thus is outside of the scope for the basic implementation this paper is focused on.

\subsubsection{Further Testing of Satellite Shell Arrangements}
In the scope of this paper, we analyze the effect changing the arrangement of LISLs aboard LEO satellites has on the network's throughput. While this knowledge is important for the developers and providers of these networks, it offers little to no benefit to users who are less concerned with comparing LISL arrangements and more concerned with the arrangement of satellites within each shell. More thorough testing is needed to determine the effect changing the number of satellites within a given shell has on the network's throughput. \par
Notably, the removal of LISLs from the entire network will likely slow down the network as the total bitrate is decreased. Rearranging those LISLs while tuning the quantity of satellites within each shell may uncover more efficient methods of populating a given satellite shell, albeit with the reduction of redundancy and thus the increase in likelihood of a single failure having wider consequences throughout the network. This research could satisfy concerns from the wider scientific community on the ecological effects that these satellite constellations produce.

\subsubsection{More Realistic Traffic Data}

In the current iteration of SLASh, traffic is generated procedurally through a rough geographic estimation of internet utilization and is limited in the types of traffic that it generates (i.e. only generates traffic flows within a particular size range). To more accurately reflect real-world conditions, it would be ideal if SLASh had a more realistic traffic model. There are two possibilities for how the traffic model could be improved: 1) Implement population demographics and fine-grained internet utilization data to improve the procedural traffic generation algorithm or 2) Capture and utilize real-world internet traffic data, extrapolating on captured results to produce realistic traffic data.\par
Both of these approaches would improve the realism of SLASh's results, but the drawback to both of these approaches is simply the time and expertise required to successfully implement them into SLASh. Improving the procedural traffic generation algorithm would be the easiest option, but would require fine-grained data about internet utilization worldwide, which can be difficult to source. Doing so would also require a decision on the fidelity of the population data, as too fine-grained of information could cause undue simulation overhead. On the other hand, utilizing real-world internet data would provide the most realistic outcome for SLASh, as it would likely include more accurate data flow sizes, and potentially even allow for the simulation of streamed data (which is a sizable portion of data traveling across the internet). However, this system would be difficult to source data for and difficult to implement.

\subsubsection{Packet Level Simulation}

SLASh approaches network traffic at a high-level flow-based abstraction. For a more realistic simulation of network conditions, this could be changed to a packet-level abstraction with a minimal implementation of transport-layer protocols like TCP or UDP. This would require a fairly large development burden, but would open the door for more detailed output, including things like packet loss and retransmission costs. This could pair well with implementing collected real-world data as discussed in the previous section, since such data would likely come in the form of packet captures.

\subsubsection{Terrestrial Connection}

Finally, a further iteration of SLASh could implement actual terrestrial connection to the simulation. LEO Satellite networks like StarLink utilize ground stations for communication, as well as "bent-pipe" ground relays for satellite-to-satellite communication, both of which are absent in SLASh. For a more realistic simulation, these terrestrial connections could be established which might impact how traffic flows throughout the network.

\bibliography{reference,zhu}

@inproceedings{b1,
  author    = {H. Wu and Y. Han and M. Rajpal and Q. Zhang and J. Wang},
  title     = {{SaTE}: Low-Latency Traffic Engineering for Satellite Networks},
  booktitle = {Proceedings of the ACM SIGCOMM 2025 Conference},
  year      = {2025},
  month     = {Sep.},
  pages     = {896--916},
  publisher = {ACM},
  address   = {São Francisco Convent Coimbra Portugal},
  doi       = {10.1145/3718958.3750524}
}

@inproceedings{b2,
  author    = {Z. Lai and others},
  title     = {{StarryNet}: Empowering Researchers to Evaluate Futuristic Integrated Space and Terrestrial Networks},
  booktitle = {20th USENIX Symposium on Networked Systems Design and Implementation (NSDI 23)},
  year      = {2023},
  month     = {Apr.},
  pages     = {1309--1324},
  editor    = {M. Balakrishnan and M. Ghobadi},
  publisher = {USENIX Association},
  address   = {Boston, MA},
  url       = {https://www.usenix.org/conference/nsdi23/presentation/lai-zeqi},
  note      = {Accessed: Dec. 15, 2025}
}

@misc{b3,
  author    = {M. Kassem and N. Sastry},
  title     = {xeoverse: A Real-time Simulation Platform for Large {LEO} Satellite Mega-Constellations},
  year      = {2024},
  doi       = {10.48550/arXiv.2406.11366},
  publisher = {arXiv}
}

@misc{b4,
  author    = {Juan Luis Cano Rodríguez},
  title     = {poliastro/poliastro: poliastro 0.17.0 ({SciPy} {US} '22 edition)},
  year      = {2022},
  month     = {Jul.},
  publisher = {Zenodo},
  doi       = {10.5281/zenodo.6817189}
}

@misc{b5,
  author    = {H. Ritchie and L. Rodés-Guirao and E. Mathieu and M. Gerber and E. Ortiz-Ospina and J. Hasell and M. Roser},
  title     = {Population by World Region},
  howpublished = {Our World in Data},
  url       = {https://archive.ourworldindata.org/20250929-103213/grapher/population-regions-with-projections.html}
}

@misc{b6,
  author    = {H. Ritchie and L. Rodés-Guirao and E. Mathieu and M. Gerber and E. Ortiz-Ospina and J. Hasell and M. Roser},
  title     = {Number of people using the Internet},
  howpublished = {Our World in Data},
  url       = {https://ourworldindata.org/grapher/number-of-internet-users}
}

@misc{b7,
  title     = {United States Government Accountability Office: Large Constellations of Satellites Mitigating Environmental and Other Effects},
  year      = {2022},
  month     = {Sep.},
  url       = {https://www.gao.gov/assets/gao-22-105166.pdf},
  note      = {Accessed: Dec. 15, 2025}
}

@misc{b8,
  title     = {Starlink},
  author    = {Wikipedia},
  year      = {2020},
  month     = {Apr.},
  url       = {https://en.wikipedia.org/wiki/Starlink}
}

@misc{b9,
  title     = {Orbital inclination},
  author    = {{Wikipedia Contributors}},
  year      = {2019},
  month     = {Sep.},
  url       = {https://en.wikipedia.org/wiki/Orbital_inclination}
}

@misc{b10,
  author    = {Siamak Hesar},
  title     = {Growing risks in low Earth orbit demand more responsible space behavior},
  howpublished = {SpaceNews},
  year      = {2024},
  month     = {Dec.},
  url       = {https://spacenews.com/growing-risks-in-low-earth-orbit-demand-more-responsible-space-behavior/}
}

@misc{b11,
  author    = {G. Rannard},
  title     = {Elon Musk’s Starlink satellites ‘blocking’ view of the universe},
  howpublished = {Bbc.com},
  year      = {2024},
  month     = {Sep.},
  url       = {https://www.bbc.com/news/articles/cy4dnr8zemgo}
}

@article{b12,
  author    = {A. U. Chaudhry and H. Yanikomeroglu},
  title     = {Laser Intersatellite Links in a Starlink Constellation: A Classification and Analysis},
  journal   = {IEEE Vehicular Technology Magazine},
  year      = {2021},
  month     = {Jun.},
  pages     = {48--56},
  doi       = {10.1109/MVT.2021.3063706}
}

@misc{b13,
  author    = {Starlink},
  title     = {Starlink | Technology},
  year      = {2025},
  url       = {https://starlink.com/technology}
}

@inproceedings{b14,
  author    = {X. Cao and X. Zhang},
  title     = {{SaTCP}: Link-Layer Informed {TCP} Adaptation for Highly Dynamic {LEO} Satellite Networks},
  booktitle = {IEEE INFOCOM 2023 - IEEE Conference on Computer Communications},
  year      = {2023},
  month     = {May},
  pages     = {1--10},
  publisher = {IEEE},
  address   = {New York City, NY, USA},
  doi       = {10.1109/INFOCOM53939.2023.10228914}
}

@misc{b15,
  author    = {{SimPy Developer Community}},
  title     = {SimPy: Discrete Event Simulation for Python},
  howpublished = {SimPy Documentation},
  url       = {https://simpy.readthedocs.io/en/latest/index.html}
}

@misc{b16,
  title     = {Ground station},
  author    = {Wikipedia},
  year      = {2022},
  month     = {Apr.},
  url       = {https://en.wikipedia.org/wiki/Ground_station}
}

@inproceedings{b17,
  author    = {A. Hagberg and D. Schult and P. Swart},
  title     = {Exploring Network Structure, Dynamics, and Function using {NetworkX}},
  booktitle = {Proceedings of the 7th Python in Science conference (SciPy 2008)},
  year      = {2008},
  pages     = {11--15},
  editor    = {G. Varoquaux and T. Vaught and J. Millman}
}

@misc{b18,
  author    = {S. Evans},
  title     = {An Intern’s Perspective: Starlink’s Megaconstellation | {RPA} Engineering},
  year      = {2024},
  month     = {Jul.},
  url       = {https://rpaengr.com/an-interns-perspective-starlinks-megaconstellation/}
}

@INPROCEEDINGS{9894326,
  author={Wei, Xiaohui and Wei, Xiukun and Wang, Xingwang and Wang, Yundi and Niu, Yan},
  booktitle={2022 IEEE International Performance, Computing, and Communications Conference (IPCCC)}, 
  title={HRCache: Edge-End Collaboration for Mobile Deep Vision Based on H.264 and Approximated Reuse}, 
  year={2022},
  volume={},
  number={},
  pages={380-388},
  doi={10.1109/IPCCC55026.2022.9894326}}

@misc{wei2025,
      title={Self-Consuming Generative Models with Adversarially Curated Data}, 
      author={Xiukun Wei and Xueru Zhang},
      year={2025},
      eprint={2505.09768},
      archivePrefix={arXiv},
      primaryClass={cs.LG},
      url={https://arxiv.org/abs/2505.09768}, 
}

@inproceedings{wire1,
author = {Wang, Wei and Liu, Xin and Chi, Zicheng and Ray, Stuart and Zhu, Ting},
title = {Key Establishment for Secure Asymmetric Cross-Technology Communication},
year = {2024},
isbn = {9798400704826},
publisher = {Association for Computing Machinery},
address = {New York, NY, USA},
url = {https://doi-org.proxy.lib.ohio-state.edu/10.1145/3634737.3637670},
doi = {10.1145/3634737.3637670},
booktitle = {Proceedings of the 19th ACM Asia Conference on Computer and Communications Security},
pages = {412–422},
numpages = {11},
location = {Singapore, Singapore},
series = {ASIA CCS '24}
}

@INPROCEEDINGS{10017581,
  author={Wang, Wei and Chi, Zicheng and Liu, Xin and Bhaskar, Ananth Vishnu and Baingane, Ankit and Jahnige, Ryan and Zhang, Qingquan and Zhu, Ting},
  booktitle={MILCOM 2022 - 2022 IEEE Military Communications Conference (MILCOM)}, 
  title={A Secured Protocol for IoT Devices in Tactical Networks}, 
  year={2022},
  volume={},
  number={},
  pages={43-48},
  doi={10.1109/MILCOM55135.2022.10017581}}

@inproceedings{10.1145/3460120.3484766,
author = {Wang, Wei and Yao, Yao and Liu, Xin and Li, Xiang and Hao, Pei and Zhu, Ting},
title = {I Can See the Light: Attacks on Autonomous Vehicles Using Invisible Lights},
year = {2021},
isbn = {9781450384544},
publisher = {Association for Computing Machinery},
address = {New York, NY, USA},
url = {https://doi-org.proxy.lib.ohio-state.edu/10.1145/3460120.3484766},
doi = {10.1145/3460120.3484766},
booktitle = {Proceedings of the 2021 ACM SIGSAC Conference on Computer and Communications Security},
pages = {1930–1944},
numpages = {15},
location = {Virtual Event, Republic of Korea},
series = {CCS '21}
}

@ARTICLE{9523755,
  author={Cheng, Long and Kong, Linghe and Gu, Yu and Niu, Jianwei and Zhu, Ting and Liu, Cong and Mumtaz, Shahid and He, Tian},
  journal={IEEE Transactions on Wireless Communications}, 
  title={Collision-Free Dynamic Convergecast in Low-Duty-Cycle Wireless Sensor Networks}, 
  year={2022},
  volume={21},
  number={3},
  pages={1665-1680},
  doi={10.1109/TWC.2021.3105983}}

@ARTICLE{9340574,
  author={Chi, Zicheng and Li, Yan and Sun, Hongyu and Huang, Zhichuan and Zhu, Ting},
  journal={IEEE/ACM Transactions on Networking}, 
  title={Simultaneous Bi-Directional Communications and Data Forwarding Using a Single ZigBee Data Stream}, 
  year={2021},
  volume={29},
  number={2},
  pages={821-833},
  doi={10.1109/TNET.2021.3054339}}

@inproceedings{10.1145/3387514.3405861,
author = {Chi, Zicheng and Liu, Xin and Wang, Wei and Yao, Yao and Zhu, Ting},
title = {Leveraging Ambient LTE Traffic for Ubiquitous Passive Communication},
year = {2020},
isbn = {9781450379557},
publisher = {Association for Computing Machinery},
address = {New York, NY, USA},
url = {https://doi-org.proxy.lib.ohio-state.edu/10.1145/3387514.3405861},
doi = {10.1145/3387514.3405861},
booktitle = {Proceedings of the Annual Conference of the ACM Special Interest Group on Data Communication on the Applications, Technologies, Architectures, and Protocols for Computer Communication},
pages = {172–185},
numpages = {14},
location = {Virtual Event, USA},
series = {SIGCOMM '20}
}

@ARTICLE{9141221,
  author={Pan, Yan and Li, Shining and Li, Bingqi and Zhang, Yu and Yang, Zhe and Guo, Bin and Zhu, Ting},
  journal={IEEE Communications Magazine}, 
  title={CDD: Coordinating Data Dissemination in Heterogeneous IoT Networks}, 
  year={2020},
  volume={58},
  number={6},
  pages={84-89},
  doi={10.1109/MCOM.001.1900473}}

@INPROCEEDINGS{9120764,
  author={Tao, Yinrong and Xiao, Sheng and Hao, Bin and Zhang, Qingquan and Zhu, Ting and Chen, Zhuo},
  booktitle={2020 IEEE Wireless Communications and Networking Conference (WCNC)}, 
  title={WiRE: Security Bootstrapping for Wireless Device-to-Device Communication}, 
  year={2020},
  volume={},
  number={},
  pages={1-7},
  doi={10.1109/WCNC45663.2020.9120764}}

@inproceedings{10.1145/3356250.3360046,
author = {Chi, Zicheng and Li, Yan and Liu, Xin and Yao, Yao and Zhang, Yanchao and Zhu, Ting},
title = {Parallel inclusive communication for connecting heterogeneous IoT devices at the edge},
year = {2019},
isbn = {9781450369503},
publisher = {Association for Computing Machinery},
address = {New York, NY, USA},
url = {https://doi-org.proxy.lib.ohio-state.edu/10.1145/3356250.3360046},
doi = {10.1145/3356250.3360046},
booktitle = {Proceedings of the 17th Conference on Embedded Networked Sensor Systems},
pages = {205–218},
numpages = {14},
location = {New York, New York},
series = {SenSys '19}
}

@INPROCEEDINGS{8737525,
  author={Wang, Wei and Liu, Xin and Yao, Yao and Pan, Yan and Chi, Zicheng and Zhu, Ting},
  booktitle={IEEE INFOCOM 2019 - IEEE Conference on Computer Communications}, 
  title={CRF: Coexistent Routing and Flooding using WiFi Packets in Heterogeneous IoT Networks}, 
  year={2019},
  volume={},
  number={},
  pages={19-27},
  doi={10.1109/INFOCOM.2019.8737525}}

@ARTICLE{8694952,
  author={Chi, Zicheng and Li, Yan and Sun, Hongyu and Yao, Yao and Zhu, Ting},
  journal={IEEE/ACM Transactions on Networking}, 
  title={Concurrent Cross-Technology Communication Among Heterogeneous IoT Devices}, 
  year={2019},
  volume={27},
  number={3},
  pages={932-947},
  doi={10.1109/TNET.2019.2908754}}

@inproceedings{10.1145/3274783.3274846,
author = {Li, Yan and Chi, Zicheng and Liu, Xin and Zhu, Ting},
title = {Passive-ZigBee: Enabling ZigBee Communication in IoT Networks with 1000X+ Less Power Consumption},
year = {2018},
isbn = {9781450359528},
publisher = {Association for Computing Machinery},
address = {New York, NY, USA},
url = {https://doi-org.proxy.lib.ohio-state.edu/10.1145/3274783.3274846},
doi = {10.1145/3274783.3274846},
booktitle = {Proceedings of the 16th ACM Conference on Embedded Networked Sensor Systems},
pages = {159–171},
numpages = {13},
location = {Shenzhen, China},
series = {SenSys '18}
}

@inproceedings{10.1145/3210240.3210346,
author = {Li, Yan and Chi, Zicheng and Liu, Xin and Zhu, Ting},
title = {Chiron: Concurrent High Throughput Communication for IoT Devices},
year = {2018},
isbn = {9781450357203},
publisher = {Association for Computing Machinery},
address = {New York, NY, USA},
url = {https://doi-org.proxy.lib.ohio-state.edu/10.1145/3210240.3210346},
doi = {10.1145/3210240.3210346},
booktitle = {Proceedings of the 16th Annual International Conference on Mobile Systems, Applications, and Services},
pages = {204–216},
numpages = {13},
location = {Munich, Germany},
series = {MobiSys '18}
}

@INPROCEEDINGS{8486349,
  author={Wang, Wei and Xie, Tiantian and Liu, Xin and Zhu, Ting},
  booktitle={IEEE INFOCOM 2018 - IEEE Conference on Computer Communications}, 
  title={ECT: Exploiting Cross-Technology Concurrent Transmission for Reducing Packet Delivery Delay in IoT Networks}, 
  year={2018},
  volume={},
  number={},
  pages={369-377},
  doi={10.1109/INFOCOM.2018.8486349}}

@inproceedings {290983,
author = {Xin Liu and Wei Wang and Guanqun Song and Ting Zhu},
title = {{LightThief}: Your Optical Communication Information is Stolen behind the Wall},
booktitle = {32nd USENIX Security Symposium (USENIX Security 23)},
year = {2023},
isbn = {978-1-939133-37-3},
address = {Anaheim, CA},
pages = {5325--5339},
url = {https://www.usenix.org/conference/usenixsecurity23/presentation/liu-xin},
publisher = {USENIX Association},
month = aug
}

@misc{song2022mlbasedsecurelowpowercommunication,
      title={ML-based Secure Low-Power Communication in Adversarial Contexts}, 
      author={Guanqun Song and Ting Zhu},
      year={2022},
      eprint={2212.13689},
      archivePrefix={arXiv},
      primaryClass={cs.CR},
      url={https://arxiv.org/abs/2212.13689}, 
}

@inproceedings{10.1145/3769102.3770620,
author = {Song, Guanqun and Li, Yan and Zhu, Ting},
title = {A Metal Sensing and Biometric-based Tracking System},
year = {2025},
isbn = {9798400722387},
publisher = {Association for Computing Machinery},
address = {New York, NY, USA},
url = {https://doi-org.proxy.lib.ohio-state.edu/10.1145/3769102.3770620},
doi = {10.1145/3769102.3770620},
booktitle = {Proceedings of the Tenth ACM/IEEE Symposium on Edge Computing},
articleno = {23},
numpages = {17},
location = {the Hilton Arlington National Landing, Arlington, VA, USA},
series = {SEC '25}
}

@misc{gopal2022securityprivacychallengesmicroservices,
      title={Security, Privacy and Challenges in Microservices Architecture and Cloud Computing- Survey}, 
      author={Hemanth Gopal and Guanqun Song and Ting Zhu},
      year={2022},
      eprint={2212.14422},
      archivePrefix={arXiv},
      primaryClass={cs.CR},
      url={https://arxiv.org/abs/2212.14422}, 
}

@misc{khatri2022heterogeneouscomputingsystems,
      title={Heterogeneous Computing Systems}, 
      author={Dimple P. Khatri and Guanqun Song and Ting Zhu},
      year={2022},
      eprint={2212.14418},
      archivePrefix={arXiv},
      primaryClass={eess.SY},
      url={https://arxiv.org/abs/2212.14418}, 
}

@misc{ketha2025analysissecurityoslevelvirtualization,
      title={Analysis of Security in OS-Level Virtualization}, 
      author={Krishna Sai Ketha and Guanqun Song and Ting Zhu},
      year={2025},
      eprint={2501.01334},
      archivePrefix={arXiv},
      primaryClass={cs.CR},
      url={https://arxiv.org/abs/2501.01334}, 
}

@misc{shergill2024energyefficientlorawanleo,
      title={Energy Efficient LoRaWAN in LEO Satellites}, 
      author={Muskan Shergill and Zach Thompson and Guanqun Song and Ting Zhu},
      year={2024},
      eprint={2412.20660},
      archivePrefix={arXiv},
      primaryClass={cs.ET},
      url={https://arxiv.org/abs/2412.20660}, 
}

@misc{ali2023security5gnetworks,
      title={Security in 5G Networks -- How 5G networks help Mitigate Location Tracking Vulnerability}, 
      author={Abshir Ali and Guanqun Song and Ting Zhu},
      year={2023},
      eprint={2312.16200},
      archivePrefix={arXiv},
      primaryClass={cs.CR},
      url={https://arxiv.org/abs/2312.16200}, 
}

@misc{yuan2024heatsatellitesmeatgpus,
      title={Heat: Satellite's meat is GPU's poison}, 
      author={Zhehu Yuan and Jinyang Liu and Guanqun Song and Ting Zhu},
      year={2024},
      eprint={2501.14757},
      archivePrefix={arXiv},
      primaryClass={cs.DC},
      url={https://arxiv.org/abs/2501.14757}, 
}

@misc{dixit2023dataclassificationmultiprocessing,
      title={Data Classification With Multiprocessing}, 
      author={Anuja Dixit and Shreya Byreddy and Guanqun Song and Ting Zhu},
      year={2023},
      eprint={2312.15152},
      archivePrefix={arXiv},
      primaryClass={cs.LG},
      url={https://arxiv.org/abs/2312.15152}, 
}

@misc{qiu2023mapreducemultiprocessinglargedata,
      title={Map-Reduce for Multiprocessing Large Data and Multi-threading for Data Scraping}, 
      author={Zefeng Qiu and Prashanth Umapathy and Qingquan Zhang and Guanqun Song and Ting Zhu},
      year={2023},
      eprint={2312.15158},
      archivePrefix={arXiv},
      primaryClass={math.NA},
      url={https://arxiv.org/abs/2312.15158}, 
}

@INPROCEEDINGS{8117550,
  author={Chi, Zicheng and Li, Yan and Yao, Yao and Zhu, Ting},
  booktitle={2017 IEEE 25th International Conference on Network Protocols (ICNP)}, 
  title={PMC: Parallel multi-protocol communication to heterogeneous IoT radios within a single WiFi channel}, 
  year={2017},
  volume={},
  number={},
  pages={1-10},
  doi={10.1109/ICNP.2017.8117550}}

@INPROCEEDINGS{8057109,
  author={Chi, Zicheng and Huang, Zhichuan and Yao, Yao and Xie, Tiantian and Sun, Hongyu and Zhu, Ting},
  booktitle={IEEE INFOCOM 2017 - IEEE Conference on Computer Communications}, 
  title={EMF: Embedding multiple flows of information in existing traffic for concurrent communication among heterogeneous IoT devices}, 
  year={2017},
  volume={},
  number={},
  pages={1-9},
  doi={10.1109/INFOCOM.2017.8057109}}

@INPROCEEDINGS{9709070,
  author={Wang, Wei and Ning, Zeyu and Iradukunda, Hugues and Zhang, Qingquan and Zhu, Ting and Yi, Ping},
  booktitle={2021 IEEE/ACM Symposium on Edge Computing (SEC)}, 
  title={MailLeak: Obfuscation-Robust Character Extraction Using Transfer Learning}, 
  year={2021},
  volume={},
  number={},
  pages={459-464},
  doi={10.1145/3453142.3491421}}

@ARTICLE{9444204,
  author={Han, Dianqi and Li, Ang and Zhang, Lili and Zhang, Yan and Li, Jiawei and Li, Tao and Zhu, Ting and Zhang, Yanchao},
  journal={IEEE/ACM Transactions on Networking}, 
  title={Deep Learning-Guided Jamming for Cross-Technology Wireless Networks: Attack and Defense}, 
  year={2021},
  volume={29},
  number={5},
  pages={1922-1932},
  doi={10.1109/TNET.2021.3082839}}

@misc{ning2021benchmarkingmachinelearningfast,
      title={Benchmarking Machine Learning: How Fast Can Your Algorithms Go?}, 
      author={Zeyu Ning and Hugues Nelson Iradukunda and Qingquan Zhang and Ting Zhu},
      year={2021},
      eprint={2101.03219},
      archivePrefix={arXiv},
      primaryClass={cs.LG},
      url={https://arxiv.org/abs/2101.03219}, 
}

@ARTICLE{8832180,
  author={Xia, Zhiyang and Yi, Ping and Liu, Yunyu and Jiang, Bo and Wang, Wei and Zhu, Ting},
  journal={IEEE Transactions on Multimedia}, 
  title={GENPass: A Multi-Source Deep Learning Model for Password Guessing}, 
  year={2020},
  volume={22},
  number={5},
  pages={1323-1332},
  doi={10.1109/TMM.2019.2940877}}

@INPROCEEDINGS{8556807,
  author={Meng, Yishuang and Yi, Ping and Guo, Xuejun and Gu, Wen and Liu, Xin and Wang, Wei and Zhu, Ting},
  booktitle={2018 Third International Conference on Security of Smart Cities, Industrial Control System and Communications (SSIC)}, 
  title={Detection for Pulmonary Nodules using RGB Channel Superposition Method in Deep Learning Framework}, 
  year={2018},
  volume={},
  number={},
  pages={1-8},
  doi={10.1109/SSIC.2018.8556807}}

@INPROCEEDINGS{8422243,
  author={Liu, Yunyu and Xia, Zhiyang and Yi, Ping and Yao, Yao and Xie, Tiantian and Wang, Wei and Zhu, Ting},
  booktitle={2018 IEEE International Conference on Communications (ICC)}, 
  title={GENPass: A General Deep Learning Model for Password Guessing with PCFG Rules and Adversarial Generation}, 
  year={2018},
  volume={},
  number={},
  pages={1-6},
  doi={10.1109/ICC.2018.8422243}}

@ARTICLE{10189210,
  author={Liu, Xin and Chi, Zicheng and Wang, Wei and Yao, Yao and Hao, Pei and Zhu, Ting},
  journal={IEEE/ACM Transactions on Networking}, 
  title={High-Granularity Modulation for OFDM Backscatter}, 
  year={2024},
  volume={32},
  number={1},
  pages={338-351},
  doi={10.1109/TNET.2023.3286880}}

@misc{chandrasekaran2022computervisionbasedparking,
      title={Computer Vision Based Parking Optimization System}, 
      author={Siddharth Chandrasekaran and Jeffrey Matthew Reginald and Wei Wang and Ting Zhu},
      year={2022},
      eprint={2201.00095},
      archivePrefix={arXiv},
      primaryClass={cs.CV},
      url={https://arxiv.org/abs/2201.00095}, 
}

@misc{iqbal2021machinelearningartificialintelligence,
      title={Machine Learning and Artificial Intelligence in Next-Generation Wireless Network}, 
      author={Wafeeq Iqbal and Wei Wang and Ting Zhu},
      year={2021},
      eprint={2202.01690},
      archivePrefix={arXiv},
      primaryClass={cs.NI},
      url={https://arxiv.org/abs/2202.01690}, 
}

@article{pan2020endogenous,
  title={Endogenous Security Defense against Deductive Attack: When Artificial Intelligence Meets Active Defense for Online Service},
  author={Pan, Yan and Li, Shining and Li, Bingqi and Zhang, Yu and Yang, Zhe and Guo, Bin and Zhu, Ting},
  journal={IEEE Communications Magazine},
  volume={58},
  number={6},
  pages={84--89},
  year={2020},
  publisher={IEEE-INST ELECTRICAL ELECTRONICS ENGINEERS INC 445 HOES LANE, PISCATAWAY, NJ~…}
}
\bibliographystyle{ieeetr}

\vspace{12pt}

\end{document}